\documentclass[aps,prd,groupedaddress,showpacs]{revtex4}
\usepackage{amsmath}
\usepackage{amssymb}
\usepackage{bbm}
\usepackage{epsfig}
\begin{document}
\begin{flushright}
SHEP-09-07
\end{flushright}
\title{Impact of FCNC top quark interactions on $BR(t\,\rightarrow\,b\,W)$}

\author{P.M. Ferreira$^{1,2}$~\footnote{ferreira@cii.fc.ul.pt} and R.
Santos$^{3}$~\footnote{rsantos@cii.fc.ul.pt}} \affiliation{$^1$
Instituto Superior de Engenharia de Lisboa, Rua Conselheiro
Em\'{\i}dio
Navarro, 1, 1959-007 Lisboa, Portugal \\
$^{2}$ Centro de F\'{\i}sica Te\'orica e Computacional, Faculdade de
Ci\^encias, Universidade de Lisboa,\\ Avenida Professor Gama Pinto,
2, 1649-003
Lisboa, Portugal; \\
$^{3}$ NExT Institute and School of Physics and Astronomy,
University of Southampton Highfield, Southampton SO17 1BJ, United
Kingdom}

\date{\today}

\begin{abstract} We study the effect that FCNC interactions of the
top quark will have on the branching ratio of charged decays of the
top quark. We have performed an integrated analysis using Tevatron and B-factories data
and with just the further assumption that the CKM matrix is unitary we can
obtain very restrictive bounds on the strong and electroweak FCNC branching
ratios $Br(t\,\rightarrow\,q\,X)\;<\; 4.0\times 10^{-4}$, where $X$ is any
vector boson and a sum in $q=u,c$ is implied.
\end{abstract}


\maketitle

\section{Introduction}

With the large statistics of top quark production expected at the
LHC - both in the $t\bar{t}$ and single top channels - precision
studies of the properties of this particle will become possible. The
detailed study of the branching ratios of the decays of the top is
of great interest, since those observables may change by several
orders of magnitude, depending on which model we are considering:
the Standard Model (SM) or some of its extensions, such as
supersymmetry (SUSY) or the two-higgs doublet models (2HDM). For
instance, the branching ratio for the flavour-changing neutral
current (FCNC) decay $t\,\rightarrow\,c\,g$ is expected to be of the
order of $10^{-12}$ in the SM, whereas in some realizations of SUSY
models it may reach as much as
$10^{-4}$~\cite{AguilarSaavedra:2004wm, calc}. Thus, a measurement
of such branching ratios may be an excellent way of discovering new
physics beyond that of the SM.

Recently, in a series of works~\cite{nos1,nos2}, we have studied the
FCNC interactions of the top quark, using the effective operator
formalism of Buchm\"uller and Wyler~\cite{buch}. That formalism
allows us to parameterize whatever new physics may exist beyond the
SM in a model-independent manner, by using operators of dimension
larger than four which obey the gauge symmetries of the SM. That is,
indeed, the greatest advantage of adopting this formalism: gauge
invariance is automatically guaranteed. In refs.~\cite{nos1}, only
operators which induced FCNC in the strong interactions of the top
quark were considered. They manifested themselves in FCNC decays of
the form $t\,\rightarrow\,u\,g$ and $t\,\rightarrow\,c\,g$, and
their impact on processes of top production at the LHC was studied
in detail. It was shown that these FCNC interactions might have
considerable impact on processes of single top production and
associated production of a top quark plus a photon, a Z boson or a
Higgs scalar. In ref.~\cite{nos2} we analysed in detail other
sources of FCNC processes, namely in the electroweak sector, with
new possible decays such as $t\,\rightarrow\,u\,\gamma$ or
$t\,\rightarrow\,c\,Z$. We showed that the top electroweak FCNC
interactions could contribute as much as the strong ones for
processes such as $t\,+\,Z$ production at the LHC. We also discussed
the possibility of experimentally distinguishing both types of FCNC
contributions. An analysis of the effect of all of these FCNC
interactions in the production of $t\,+$ jet at the LHC has recently
been concluded~\cite{nos3}.

In this work we wish to study the contributions that these FCNC
interactions may have on a more basic top observable: the branching
ratio of $t\,\rightarrow\,b\,W$, which is expected, in the SM, to be
very close to 1. An obvious way in which the top FCNC interactions
will affect $BR(t\,\rightarrow\,b\,W)$ is by changing the total top
width: since
$BR(t\,\rightarrow\,b\,W)\,=\,\Gamma(t\,\rightarrow\,b\,W)/\Gamma_t(total)$,
with anomalous FCNC decays allowed the value of the total top width
will surely increase, thus, in principle, decreasing the value of
$BR(t\,\rightarrow\,b\,W)$. However, we will show in this paper the
existence of a more curious effect, namely, that gauge invariance
implies that the same operators which generate FCNC top decays also
give contributions to charged top decays of the form
$t\,\rightarrow\,d\,W$ and $t\,\rightarrow\,s\,W$.

This paper is organized as follows: in section~\ref{sec:op} we will
briefly review the effective operator formalism and the criteria
behind the selection of those operators for FCNC purposes. We will
also review the results of~\cite{nos1,nos2} concerning the FCNC
decay widths of the top quark. In section~\ref{sec:br} we will
present the contributions of the FCNC operators to the charged top
decays, and the calculation of the branching ratio for
$t\,\rightarrow\,b\,W$. In section~\ref{sec:calc} we will show
predictions for the FCNC cross sections and branching ratios in
terms of $BR(t\,\rightarrow\,b\,W)$ and discuss the importance that
precision measurements of this observable might have on the cross
section of processes such as $t\,+\,Z$ and $t + jet$ production at
the LHC. An overview of our conclusions is presented in
section~\ref{sec:conc}.

\section{Effective operators contributing to top FCNC interactions}
\label{sec:op}

The effective operator formalism~\cite{buch} considers that,
whatever new physics may exist beyond the SM, it will reflect itself
at low energies as operators of dimension larger than four, built
with the same fields of the SM and respecting the gauge symmetries
$SU(3)_C \times SU(2)_W \times U(1)_Y$. The SM is thus seen as the
low-energy limit of a more general theory, the effects of which are
described in a model-independent manner by operators of dimension 5,
6, etc. Hence, the lagrangian of the theory becomes
\begin{equation}
{\cal L} \;\;=\;\; {\cal L}^{SM} \;+\; \frac{1}{\Lambda}\,{\cal
L}^{(5)} \;+\; \frac{1}{\Lambda^2}\,{\cal L}^{(6)} \;+\;
O\,\left(\frac{1}{\Lambda^3}\right) \;\;\; , \label{eq:l}
\end{equation}
where ${\cal L}^{SM}$ is the usual SM lagrangian, ${\cal L}^{(5)}$
and ${\cal L}^{(6)}$ are the dimension five and six lagrangians and
$\Lambda$ is the energy scale for which one expects that whatever
new physics exists beyond the SM becomes relevant. For our purposes
- studies at the LHC - $\Lambda$ will be of the order of one TeV. Note however that we will vary the coupling constants, denoted generically as $|a/\Lambda^2|$, from $10^{-12}$ $TeV^{-2}$ to 1 $TeV^{-2}$ although some of the constraints from B physics will immediately discard some of values generated.
This means that for an $a$ of order 1, $\Lambda$ is allowed to vary between 1 $TeV$ and $10^6$ $TeV$ - 1 $TeV$ just sets the scale of energy.
The contributions from ${\cal L}^{(5)}$ break baryon and lepton
number, and do not contribute to the FCNC physics we are interested
in studying. The number of operators contained in ${\cal L}^{(6)}$ is enormous. Due to the gauge structure of the SM, a given dimension 6 operator that has an impact on top interactions can also have a parallel effect on processes involving only bottom quarks. B physics~\cite{Fox:2007in} is obviously the main source of constraints. The analysis of~\cite{Fox:2007in} as well as the one in~\cite{Grzadkowski:2008mf} (see also~\cite{other1}) can be used to impose limits on the sizes of several of our anomalous couplings, due to experimental constraints from B physics like $b \to s \gamma$ and others.
There is a hierarchy of constraints resulting from the underlying SM gauge structure. The set of operators where the gauge structure manifests more strongly is the one denoted by $LL$ in~\cite{Fox:2007in} as these operators are built with only $SU(2)$ doublets. Operators $RR$, built with singlets alone, are obviously the least constrained as no relation exists between an $R$-top and a $R$-bottom.
The set of constraints obtained in~\cite{Fox:2007in} shows that operators of the type $LL$ are already constrained beyond the reach of the LHC. Using the LHC~\cite{toni, fla, CMS} predictions they could show that some of the constraints on the operators coming from B physics are already stronger than what is predicted that could be measured at the LHC. This is true for operators of type $LL$, while limits on $LR$ and $RL$ operators are close to what is expected to be measured at the LHC. Hence, we discard all $LL$ operators and we will use all the constraints on the dimension-6 operators obtained in~\cite{Fox:2007in}. B factories and the Tevatron are still collecting data and therefore these constraints will be even stronger by the time the LHC starts to analyse data. Finally we should mention that there are specific models for which some, but not all, effects on B physics of operators of type $LL$ built with doublets only, cancel exactly~\cite{mpv}. In the model presented in that paper, anomalous contributions to vertices involving the $Z$ boson and $b$ quarks cancel precisely. However, the same does not occur for vertices with the $W$ boson and $b$ quarks - if we wish to cancel those contributions, we need to adjust the values of some of the anomalous couplings to make it happen. We still choose to leave those operators out of our analysis, but their inclusion may be of interest in the study of very specific models. Finally we note that B physics only constraints operators with origin in the electroweak sector.

The criteria mentioned above are very strict, and the number of
operators which survive the selection is small. For the strong
sector, we have the operators~\cite{nos1}
\begin{align}
{\cal O}_{itG} &= i \frac{\alpha^S_{it}}{\Lambda^2}\,
\left(\bar{u}^i_R \, \lambda^{a} \, \gamma_{\mu}  D_{\nu} t_R\right)
\, G^{a \mu \nu} \;\;\;,\;\;\;{\cal
O}_{itG\phi}=\frac{\beta^{S}_{it}}{\Lambda^2}\,\left(\bar{q}^i_L\,
\, \lambda^{a} \, \sigma^{\mu\nu}\,t_R\right)\, \tilde{\phi}
\,G^{a}_{\mu\nu}\;\;\; . \label{eq:opst}
\end{align}
The coefficients $\alpha^S_{it}$ and $\beta^{S}_{it}$ are complex
dimensionless couplings, $u^i_R$ and $q^i_L$ represent the
right-handed up-type quark and left-handed quark doublet of the
first and second generation and $G^a_{\mu\nu}$ is the gluonic field
tensor. There are also operators, with couplings $\alpha^S_{ti}$ and
$\beta^{S}_{ti}$, where the positions of the top and $u^î$, $q^i$
spinors are exchanged. The lagrangian obviously also includes the
hermitian conjugates of these operators.

These operators generate FCNC vertices of the form $t\,q\,g$ or
$t\,q\,g\,g$, among others (where $q\,=\,u\,,\,c$). The constants
$\{\alpha^S\,,\,\beta^S\}$ are {\em a priori} unknown, the magnitude
of which determines the importance of the FCNC processes. The
Feynman rules for the FCNC vertices involving gluons stemming from
these operators may be found in~\cite{nos1}, and with those it is
simple to calculate the decay width of $t\,\rightarrow\,q\,g$, given
by (all quark masses except for the top's were set equal to zero)
\begin{align}
\Gamma (t \rightarrow q g) &=\;  \frac{m^3_t}{12
\pi\Lambda^4}\,\Bigg\{ m^2_t \,\left|\alpha_{tq}^S  +
(\alpha^S_{qt})^* \right|^2 \,+\, 16 \,v^2\, \left(\left|
\beta_{tq}^S \right|^2 + \left| \beta_{qt}^S \right|^2 \right)
\;\;\; +
\vspace{0.3cm} \nonumber \\
 & \hspace{2.2cm}\, 8\, v\, m_t\,\mbox{Im}\left[ (\alpha_{qt}^S  + (\alpha^S_{tq})^*)
\, \beta_{tq}^S \right] \Bigg\} \label{eq:tqg}\;\;\; .
\end{align}

The electroweak FCNC interactions of the top quark were studied
in~\cite{nos2}, where it was shown that, according to the selection
criteria described above, the operators describing them were given
by
\begin{align}
{\cal O}_{tB}= i \frac{\alpha^B_{it}}{\Lambda^2}\,\left(\bar{u}^i_R
\, \, \gamma_{\mu} D_{\nu} t_R \right) \, B^{\mu \nu}\;\;\; , &
\;\;\;{\cal O}_{tB\phi} =
\;\;\frac{\beta^{B}_{it}}{\Lambda^2}\,\left(\bar{q}^i_L\,
\sigma^{\mu\nu}\,t_R\right)\, \tilde{\phi} \,B_{\mu\nu} \;\;\; , \nonumber \\
{\cal
O}_{tW\phi}=\frac{\beta^{W}_{it}}{\Lambda^2}\,\left(\bar{q}^i_L\, \,
\tau_{I} \, \sigma^{\mu\nu}\,t_R\right)\,
\tilde{\phi}\,W^{I}_{\mu\nu}\;\;\; , & \;\;\; {\cal O}_{\phi_t}
 \, = \, \theta_{it} \, (\phi^{\dagger} D_{\mu} \phi) \, (\bar{u^i_R} \gamma^{\mu} t_R)  \;\;\;
 , \nonumber \\
 {\cal O}_{D_t} =\frac{\eta_{it}}{\Lambda^2}\,\left(\bar{q}^i_L\,
D^{\mu}\,t_R\right)\, D_{\mu} \tilde{\phi} \, \;\;\;, & \;\;\; {\cal
O}_{\bar{D}_t}=\frac{\bar{\eta}_{it}}{\Lambda^2}\,\left( D^{\mu}
\bar{q}^i_L\, \,t_R\right)\, D_{\mu} \tilde{\phi} \;\;\; .
\label{eq:opew}
\end{align}
Once more, the complex constants
$\{\alpha\,,\,\beta\,,\,\theta\,,\,\eta\,,\,\bar{\eta}\}$ are
unknown, and used to parameterize the magnitude of the FCNC
interactions.

One can use the equations of motion of the fields to establish
relations between several of the operators shown above (see, for
instance, eqs.~(8) in~\cite{nos1} and more recently, the work
of~\cite{juanm}).  With our conventions, they translate, for the
strong sector operators, as
\begin{align}
{\cal O}^{\dagger}_{utG} &= - {\cal O}_{tuG}\;+\;\frac{i}{2}
(\Gamma^{\dagger}_u\,
{\cal O}^{\dagger}_{u t G\phi} \, - \, \Gamma_t \,{\cal O}_{t u G\phi}) \nonumber \\
{\cal O}^{\dagger}_{utG} &= {\cal O}_{tuG}\; + \;i\, g_s\,
\bar{t}\, \gamma_{\mu}\, \gamma_R\, \lambda^a\,u\, \sum_i
(\bar{u}^i\, \gamma^{\mu}\, \gamma_R\, \lambda_a u^i\,+\,
\bar{d}^i\, \gamma^{\mu}\, \gamma_R\, \lambda_a\, d^i) \;\;\; .
\label{eq:re3}
\end{align}
where $\Gamma_i$ are Yukawa couplings and the conventions
of~\cite{buch} were used. The first of these equations establishes
that one can safely set to zero one of the anomalous couplings
$\{\alpha\,,\,\beta\}$. The second one specifies that, if one were
to include four-fermion operators in our calculation, one could
set one further $\{\alpha\,,\,\beta\}$ coupling to zero - or, in
exchange, to set one of the four-fermion couplings to zero. This
is precisely what we have done in~\cite{nos2}. If we were
analysing processes which only involved a top quark alongside
gauge bosons (for instance, $t\,+\,Z$ production) it is arguable
that one might then use both equations~\eqref{eq:re3} and keep
only two strong anomalous couplings, since no four-fermion terms
contribute to those processes. However, in the work of
refs.~\cite{nos1,nos3}, as well as in the current paper, we have
performed a global analysis of the impact of FCNC top
interactions. We have studied several channels simultaneously,
namely top plus gauge boson production as well as four fermion
channels (such as $q\,\bar{q}\,\rightarrow\,t\,\bar{q}$). As such,
unless we were to include all four-fermion operators in our
calculations, we are not allowed to use the second
equation~\eqref{eq:re3} to eliminate any coupling.

The importance of the simultaneous analysis of all channels is shown
in ref.~\cite{nos3}, where we computed the total anomalous cross
section for single top plus jet production at the Tevatron - and
then proceeded to use their experimental results to constrain the
values of the anomalous couplings and several branching ratios.
Further, as can be seen in the expressions presented in~\cite{nos3},
there are clearly {\em three} independent functions multiplying the
several combinations of $\{\alpha\,,\,\beta\}$ couplings. This
indicates, as discussed above, that by means of using the first
equation of~\eqref{eq:re3} we could ``drop" one of those couplings,
but not two of them. Nevertheless, we keep all
$\{\alpha\,,\,\beta\}$ couplings since no significant simplification
would come out of eliminating just one $\{\alpha\,,\,\beta\}$
coupling (none of the anomalous vertices shown in~\cite{nos1,nos3}
would be discarded). Finally, we found that this small coupling
redundancy is quite useful as a cross-check of our computations.

Due to the standard Weinberg rotation we define new constants
directly associated with the photon and the Z, given by
\begin{align}
\alpha^{\gamma}\, = \,\cos\theta_W \, \alpha^{B} \qquad \;\;\; ,&
\;\;\; \beta^{\gamma} \, = \, \sin\theta_W \beta^{W} + \cos \theta_W
\beta^{B}\;\;\; ,\nonumber \\
\alpha^{Z}\, = \, - \sin\theta_W \, \alpha^{B} \;\;\; , & \;\;\;
\beta^{Z} \, = \, \cos\theta_W \beta^{W} - \sin \theta_W \beta^{B}
  \;\;\; . \label{eq:wein}
\end{align}
In terms of these anomalous couplings the decay widths for the
decays $t\,\rightarrow\,q\,\gamma$ and $t\,\rightarrow\,q\,Z$ are
given by~\cite{nos2}
\begin{align}
\Gamma (t \rightarrow q \gamma) &=\;  \frac{m^3_t}{64
\pi\Lambda^4}\,\Bigg\{ m^2_t \,\left|\alpha_{tq}^{\gamma}  +
(\alpha^{\gamma}_{qt})^* \right|^2 \,+\, 16 \,v^2\, \left(\left|
\beta_{tq}^{\gamma} \right|^2 + \left| \beta_{qt}^{\gamma} \right|^2
\right) \;\;\; +
\vspace{0.3cm} \nonumber \\
 & \hspace{2.2cm}\, 8\, v\, m_t\,\mbox{Im}\left[ (\alpha_{qt}^{\gamma}  + (\alpha^{\gamma}_{tq})^*)
\, \beta_{tq}^{\gamma} \right] \Bigg\} \label{eq:tqga}
\end{align}
and
\begin{eqnarray}
 \Gamma(t\,\rightarrow \,q\,Z) & = & \frac{{\left( m_t^2 - m_Z^2 \right) }^2}{32\,m_t^3\,\pi
\,\Lambda^4}
\left[ K_1 \, \left| \alpha^Z_{qt} \right|^2 + K_2 \, \left|
\alpha^Z_{tq} \right|^2 + K_3 \, ( \left| \beta^Z_{qt} \right|^2 +
\left| \beta^Z_{tq} \right|^2)+ K_4 \, ( \left| \eta_{qt} \right|^2
+ \left| \bar{\eta}_{qt} \right|^2) \right.
\nonumber \\[0.25cm]
&&  \qquad + \, K_5 \, \left| \theta \right|^2 + K_6 \, Re \left[
\alpha^Z_{qt} \, \alpha^Z_{tq} \right] + K_7 \, Im \left[
\alpha^Z_{qt} \, \beta^Z_{tq} \right]
\nonumber \\[0.25cm]
&&  \qquad + \, K_8 \, Im \left[ \alpha^{Z^*}_{tq} \, \beta^Z_{tq}
\right] + K_9 \, Re \left[ \alpha^Z_{qt} \theta^* \right]+ K_{10} \,
Re \left[ \alpha^Z_{tq} \theta \right]
\nonumber \\[0.25cm]
&& \qquad   \left. + \, K_{11} \, Re \left[ \beta^Z_{qt}
(\eta_{qt}-\bar{\eta}_{qt})^* \right] + K_{12} \, Im \left[
\beta^Z_{tq} \, \theta \right] + K_{13} \, Re \left[ \eta_{qt}
\bar{\eta}_{qt}^* \right] \right]\;\;\; , \label{eq:tqZ}
\end{eqnarray}
with coefficients $K_i$ given by
\begin{eqnarray}
K_1 & = & \frac{1}{2} \, (m_t^4 + 4\,m_t^2\,m_Z^2 + m_Z^4) \qquad
K_2 \, = \, \frac{1}{2} \,  (m_t^2 - m_Z^2)^2 \qquad K_3 \, = \,
4\,( 2\,m_t^2 + m_Z^2) \,v^2
 \nonumber  \\
K_4 & = & \frac{v^2}{4\,m_Z^2}(m_t^2 - m_Z^2)^2 \qquad K_5 \, = \,
\frac{v^4}{m_Z^2} ( m_t^2 + 2\,m_Z^2 ) \qquad K_6 \, = \, ( m_t^2 -
m_Z^2 ) \,( m_t^2 + m_Z^2)
\nonumber \\
K_7 & = & 4\,m_t\, ( m_t^2 + 2\,m_Z^2 ) \,v \qquad K_8 \, = \,
4\,m_t\, ( m_t^2 - m_Z^2 ) \,v \qquad K_9 \, = \, -2\, ( 2\,m_t^2 +
m_Z^2 ) \,v^2
\nonumber \\
K_{10} & = & -2\, ( m_t^2 - m_Z^2) \,v^2 \qquad K_{11} \, = - K_{10}
\qquad K_{12} \, = \, -12 \,m_t\,v^3 \qquad K_{13} \, = \, \frac{-
v^2 }{m_Z^2} \, K_2 \;\;\;\ .
\end{eqnarray}
In expressions~\eqref{eq:tqg}, ~\eqref{eq:tqga} and~\eqref{eq:tqZ},
the couplings $\alpha^S_{qt}$, etc, correspond to the separate cases
$q\,=\,u$ and $q\,=\,c$. That is, the FCNC anomalous couplings will,
in general, have different values for the decays of the ``u" or ``c"
quarks.

\section{FCNC contributions to $BR(t\,\rightarrow\,b\,W)$}
\label{sec:br}

\begin{figure}[htbp]
  \begin{center}
    \epsfig{file=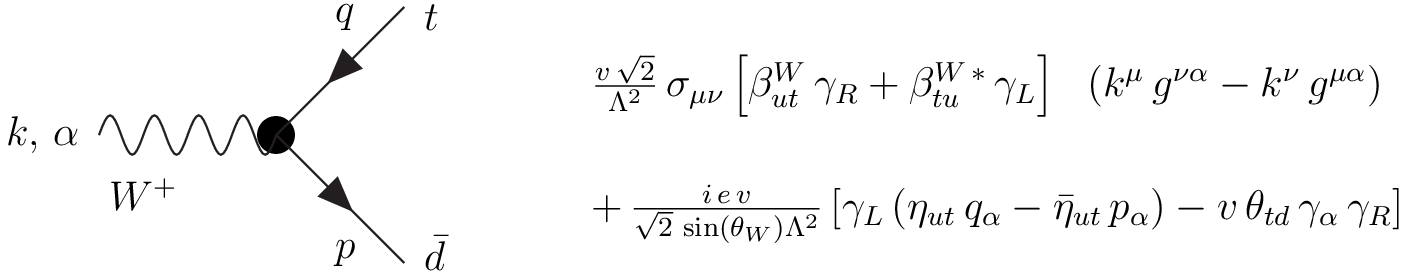,width=15 cm}
    \caption{Feynman rules corresponding to anomalous contributions to $t \,\rightarrow\, d\, W$.}
    \label{fig:feyn}
  \end{center}
\end{figure}

As we observe from eqs.~\eqref{eq:wein}, the fact that the
Buchm\"uller and Wyler formalism is automatically gauge invariant
imposes severe constraints on the type of new physics predicted by
the effective operators we are considering. Namely, those equations
show us that FCNC processes involving the photon cannot be entirely
disentangled from those involving the $Z$, since the same operators
which contribute to decays of the form $t\,\rightarrow\,q\,\gamma$
also contribute to $t\,\rightarrow\,q\,Z$. A careful analysis of the
operators in eq.~\eqref{eq:opew} shows us that some of them actually
do contribute to decays of the form $t\,\rightarrow\,d\,W$ and
$t\,\rightarrow\,s\,W$ - namely, the operators ${\cal O}_{tW\phi}$,
${\cal O}_{D_t}$ and ${\cal O}_{\bar{D}_t}$, from whose structure
one can ``extract" the correct vertices~\footnote{The absence of any
anomalous contributions to the decay $t\,\rightarrow\,b\,W$ is a
direct consequence of the physical criteria we applied to the
selection of the effective operators. Namely, the exclusion of any
operators having a direct impact on bottom quark physics.}. The
constants $\alpha$ do not contribute to these decays, as they stem
from operators which involve only right-handed quarks. Neither does
the operator ${\cal O}_{\phi_t}$, which has no $d$ spinors. However,
one can consider another operator of the type of ${\cal
O}_{\phi_t}$, namely
\begin{equation}
{\cal O}_{\phi_{td}} \, = \, i \frac{\theta_{td}}{\Lambda^2}\,
(\phi^\dagger \epsilon D_\mu \phi) \, (\bar{t}_R \, \gamma_{\mu} d_R
) \;\;\; .
\end{equation}
Gathering these operators, then, we arrive at the contribution of
the FCNC top quark effective operators on its charged decays. This
translates as a Feynman rule for ``anomalous" top W interactions,
presented in fig.~\eqref{fig:feyn}. In this figure we consider only
the vertex $W^+\,t\,\bar{d}$, there is an analogous vertex for
$W^+\,t\,\bar{s}$ involving the anomalous ``c" couplings.
One important issue needs to be considered: none of these anomalous
operators has any contribution to the masses of the quarks. Thus,
these contributions to the vertices $W\,t\,\bar{d}$ do not change
the rotation between weak eigenstates of the quarks and their mass
eigenstates. As such, these anomalous interactions constitute extra
contributions to the CKM top matrix elements.

A subtlety now affects this calculation: in refs.~\cite{nos1,nos2}
we performed calculations of observables up to order $1/\Lambda^4$.
Indeed, the FCNC decay widths of
eqs.~\eqref{eq:tqg},~\eqref{eq:tqga} and~\eqref{eq:tqZ} are of this
order. However, that was due to the fact that, at tree-level, there
are no Feynman diagrams contributing to the processes considered
($t\,\rightarrow\,q\,g$ decays, $t\,+\,Z$ production at the LHC,
etc.). For the charged decays of the top, however, there are
tree-level diagrams one must consider. Thus, the first and leading
contribution to these processes stems from the interference terms
between the SM process and the anomalous one, and is therefore of
order $1/\Lambda^2$. No contributions of order $1/\Lambda^4$ will be
considered here - these would have to include the square of the
anomalous vertex of order $1/\Lambda^2$ (which we could compute
easily) and the interference terms between the SM diagrams and those
coming from effective operators of order $1/\Lambda^4$ (which we
have not considered in this work).

Even though the anomalous vertex in fig.~\eqref{fig:feyn} includes
contributions from several FCNC couplings, only $\beta^W_{ut}$ (and
$\beta^W_{ct}$) will have an impact on the decay
$t\,\rightarrow\,d\,W$ (and $t\,\rightarrow\,s\,W$). This is due to
the fact that we are considering all quark masses other than the
top's equal to zero, and the chiral structure of the
interactions in the vertex of fig.~\eqref{fig:feyn}. A simple
calculation yields,
\begin{equation}
\Gamma (t\,\rightarrow\,d\,W)\;=\;1.42\,|V_{td}|^2\;
-\;\frac{3\,g}{8\,\pi\,\Lambda^2}\,\frac{m_t^4\,-\,m_W^4}{m_t^2}\,v\,{\mbox
Re}\left(V_{td}\,\beta^W_{ut}\right) \;\;\; , \label{eq:tdW}
\end{equation}
where the first term is the SM contribution~\cite{br}, and the
second one the interference with the anomalous operators. Thus, the
branching ratio of $t\,\rightarrow\,b\,W$ will be given by
\begin{equation}
BR(t\,\rightarrow\,b\,W)\;\;=\;\;\frac{\Gamma
(t\,\rightarrow\,b\,W)}{\sum_{q=u,c} \left[\Gamma
(t\,\rightarrow\,q\,g)\,+\,\Gamma
(t\,\rightarrow\,q\,\gamma)\,+\,\Gamma
(t\,\rightarrow\,q\,Z)\right]\,+\,\sum_{q=d,s,b} \Gamma
(t\,\rightarrow\,q\,W)} \;\;\; , \label{eq:br}
\end{equation}
where $\Gamma (t\,\rightarrow\,b\,W)\,=\,1.42\,|V_{tb}|^2$ GeV and
the remaining widths are given by
eqs.~\eqref{eq:tqg},~\eqref{eq:tqga},~\eqref{eq:tqZ}
and~\eqref{eq:tdW}. This equation tells us that {\em all} the
anomalous couplings considered so far have an impact on this
branching ratio.

\section{Analysis of results}
\label{sec:calc}

In refs.~\cite{nos1,nos2} we have shown that the top quark anomalous
FCNC interactions might have a large impact on single top production
cross sections at the LHC. Furthermore, and due to the
gauge-invariant nature of the formalism, we have shown that several
FCNC observables are correlated. We also discovered a proportionality between several
cross sections and certain FCNC widths. In this work we have shown
that the same anomalous couplings which contribute to single top
production at the LHC might also have a sizeable effect on a basic
top quark property, its largest branching ratio. Thus, to probe the
consistency of the formalism, an analysis of the expected deviations
from the SM expected value, due to FCNC interactions, is in order.

A few words on the current state of knowledge about
$Br(t\,\rightarrow\,b\,W)$: in the SM, at tree-level, this quantity
is simply given by $|V_{tb}|^2$, assuming that the CKM matrix is
unitary. Direct measurements of the top CKM coefficients are
obviously quite difficult, and the corresponding results are
affected by large error bars~\cite{pdg}. Nevertheless, the
assumption of the unitarity of the CKM matrix is an extremely
powerful one, and allows one to quote very precise
values for these numbers which include all experimental results do date plus the theoretical input of unitarity. Namely, the PDG-listed values
are~\cite{pdg}
\begin{equation}
V_{td} \,=\, 0.00874_{-0.00037}^{+0.00026}\;\;,\;\; V_{ts} \,=\,
0.0407\,\pm 0.0010\;\;,\;\;V_{tb} \,=\,
0.999133_{-0.000043}^{+0.000044}\;\;\; . \label{eq:ckm}
\end{equation}
As such, the SM prediction for $Br(t\,\rightarrow\,b\,W)$ is
extremely precise, if one assumes the unitarity of the CKM matrix.
All of our results which include FCNC interactions are therefore
compared with the SM values under this assumption, and any eventual
deviations will have to be interpreted in this light.

To investigate the importance that FCNC interactions might have on
$Br(t\,\rightarrow\,b\,W)$, we made a vast scan of the possible
values of the FCNC anomalous couplings. We considered almost two
million different sets of anomalous couplings, varying them by
several orders of magnitude, namely
$10^{-12}\,\leq\,|a/\Lambda^2|\,\leq\,1$ TeV$^{-2}$, where {\em a}
is a generic anomalous coupling. The upper limit in this variation
comes from unitarity considerations~\cite{wud}: given that the
effective operator formalism is not renormalizable, it is not
advisable to consider large values of the couplings, lest the growth
of the anomalous contributions with $\sqrt{s}$ becomes too steep. We
also demanded that all partial FCNC anomalous branching fractions
to be in agreement with all experimental bounds from
LEP~\cite{LEP2Zgamma}, Hera~\cite{Hera} and
from the Tevatron~\cite{Aaltonen:2008qr,tZqCDF,other2}.
For a detailed discussion about the referred experimental
bounds see~\cite{nos3}.

The same anomalous couplings which we use to compute FCNC branching
ratios also enter into the expressions for FCNC processes of single
top production, such as the $t\,+\,Z$ or $t\,+\,jet$ channels -
those expressions may be found in refs.~\cite{nos1,nos2,nos3}. The
cross sections were obtained using a cut on the $p_T$ of the final
state particles of 15 GeV. The partonic cross sections were
integrated with CTEQ6M parton density functions~\cite{cteq6}, by
setting the factorization scale equal to $m_t$, if the final state
is a top quark and a jet, and $m_t\,+\,m_Z$, if the final state is
$t\,+\,Z$. As we did in ref.~\cite{nos3}, we will use the existing
data from the Tevatron experiments concerning single top production
to constrain our anomalous
couplings. Namely, we will demand that, for a given set of anomalous
couplings, the value of the extra FCNC contributions to the cross
section for the single top + jet channel (which adds to its SM
value) at the Tevatron be inferior to the observed experimental
error for that observable~\cite{crosstev}, meaning 0.6 pb. This will
eliminate a significant portion of our parameter space.
We have also included all the constraints on the anomalous couplings
derived from B physics in~\cite{Fox:2007in}. These constraints are
really important as they further reduce the allowed parameter
space improving the bounds on FCNC cross sections and branching ratios.

\begin{figure}[ht]
  \begin{center}
    \epsfig{file=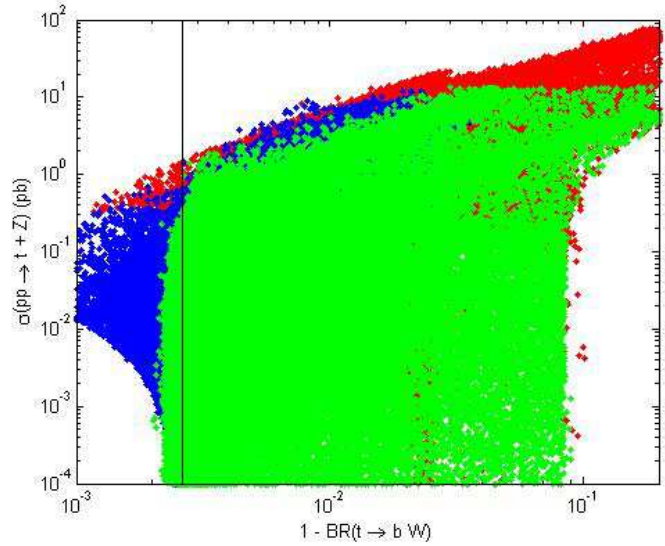,width=10 cm}
    \caption{FCNC cross section for $t\,+\,Z$ production at the LHC,
    versus $1\,-\,Br(t\,\rightarrow\,b\,W)$. The red points correspond to
    those portions of parameter space excluded by the Tevatron results,
    while the blue points are excluded by the B physics analysis in~\cite{Fox:2007in}. The 
    solid line is the SM-predicted value for $1\,-\,Br(t\,\rightarrow\,b\,W)$ and the points in green
    are those which comply with both the Tevatron and
    B-factories data.}
    \label{fig:tZbrw}
  \end{center}
\end{figure}

In figure~\ref{fig:tZbrw} we plot the values of the FCNC cross
section for $t\,+\,Z$ production at the LHC versus the values of
$1\,-\,Br(t\,\rightarrow\,b\,W)$. The central solid line corresponds
to the SM-predicted value, assuming the CKM matrix is unitary. The
error bar stemming from the values shown in eq.~\eqref{eq:ckm} is
negligible. The points in red correspond to those combinations of
anomalous couplings which were rejected due to the experimental
constraints from the Tevatron, which for this observable do not
exclude a significant region of the plot. The blue points
are the ones excluded by the B physics analysis in~\cite{Fox:2007in}.
The most important aspect
of this plot is the fact that, if one were to measure at the LHC,
with some precision, the value of the branching ratio for
$t\,\rightarrow\,b\,W$ at the LHC and find a value very close to
that which is to be expected if the SM holds, then one would expect
the cross section for $t\,+\,Z$ production at the LHC to be smaller
than about 1 pb. In fact, the expected SM value for the branching
ratio provides us an upper bound on that cross section.
Alternatively, if one were to measure a cross section for $t\,+\,Z$
production larger than approximately 1 pb, then the assumption that presides the
calculation of $Br(t\,\rightarrow\,b\,W)$ in this figure would be
incorrect - meaning, the values of the top CKM matrix elements would
be quite different from those of eq.~\eqref{eq:ckm}, and the CKM
matrix could not be considered unitary.

\begin{figure}[ht]
  \begin{center}
    \epsfig{file=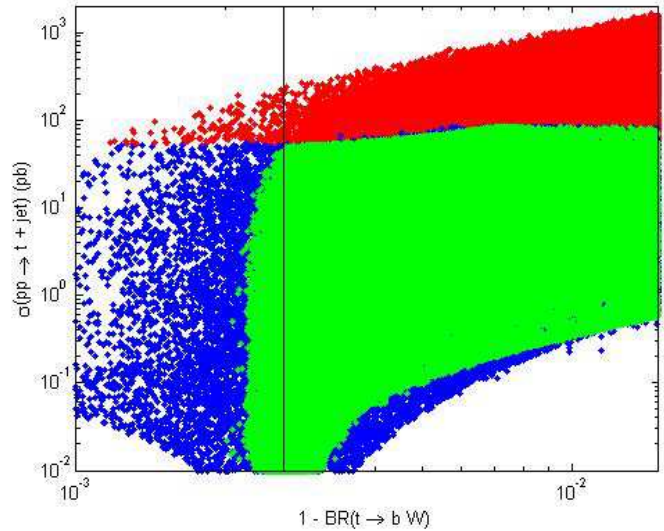,width=10 cm}
    \caption{FCNC contributions to cross section for $t\,+\,jet$ production at the LHC,
    versus $1\,-\,Br(t\,\rightarrow\,b\,W)$. The red points correspond to
    those portions of parameter space excluded by the Tevatron results,
    while the blue points are excluded by the B physics analysis in~\cite{Fox:2007in}. The
    solid line is the SM-predicted value for $1\,-\,Br(t\,\rightarrow\,b\,W)$ and the points in green
    are those which comply with both the Tevatron and
    B-factories data.}
    \label{fig:tjbrw}
  \end{center}
\end{figure}
\begin{figure}[ht]
  \begin{center}
    \epsfig{file=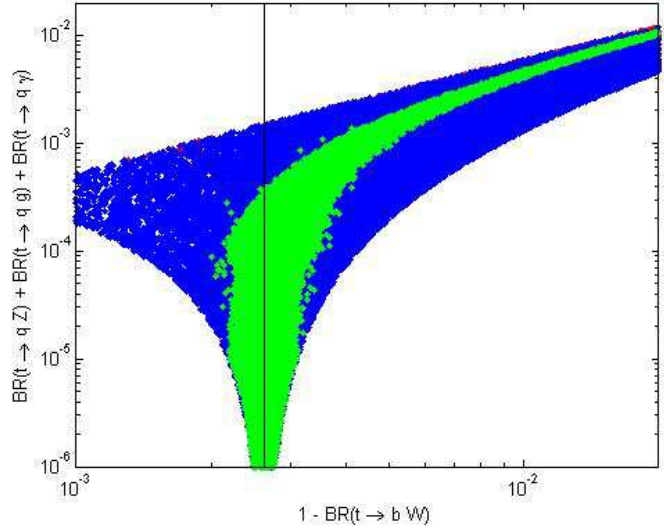,width=10 cm}
    \caption{Total FCNC branching ratios
    versus $1\,-\,Br(t\,\rightarrow\,b\,W)$. The points in green
    are those which comply with both the Tevatron and
    B-factories data. The red points correspond to
    parameter space excluded by the Tevatron and the blue points are excluded
    by the B physics analysis in~\cite{Fox:2007in}}
    \label{fig:qXbrw}
  \end{center}
\end{figure}
\begin{figure}[ht]
  \begin{center}
    \epsfig{file=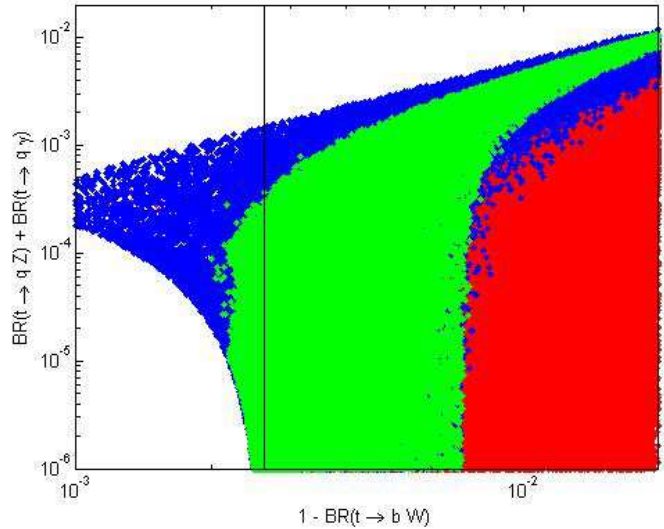,width=10 cm}
    \caption{Electroweak FCNC branching ratios
    versus $1\,-\,Br(t\,\rightarrow\,b\,W)$. The red points correspond to
    those portions of parameter space excluded by the Tevatron results,
    while the blue points are excluded by the B physics analysis in~\cite{Fox:2007in}. The
    solid line is the SM-predicted value for $1\,-\,Br(t\,\rightarrow\,b\,W)$ and the points in green
    are those which comply with both the Tevatron and
    B-factories data.}
    \label{fig:brWbrwe}
  \end{center}
\end{figure}
Consider now figure~\ref{fig:tjbrw}, where we plot the total FCNC
contribution to the top + jet cross section~\footnote{As was
explained in refs.~\cite{nos1,nos2,nos3}, the FCNC top + jet cross
section contributions add to those expected from the SM. } at the
LHC. The points which survive the Tevatron and the B-factories data are
the light, green ones, and we can see that both the Tevatron and the B-factories
exclude a significant portion of available parameter space. The line
corresponding to the SM prediction for
$1\,-\,Br(t\,\rightarrow\,b\,W)$ once again establishes an upper
bound for the FCNC contribution to this cross section, namely about
40 pb. The predicted SM value for single top production at the LHC
is around 202 pb, with a theoretical error of 8 pb~\cite{crosstev2}. This
includes the $t\,+\,W$ channel. If one considers only the $s$ and
$t$ channel contributions (so as to obtain an estimate for the SM
top + jet cross section) the value is $\sigma^{SM}
(p\,p\,\rightarrow\,t\,+\,jet)\,=\,168\,\pm\,6$ pb.
In this case the t-channel cross section is expected
to be measurable with a total error of 10 \% at the LHC
while an uncertainty of 36 \% was estimated for the s-channel
mode~\cite{crosstev3}. Therefore, the error in this quantity
will probably be inferior to the maximum FCNC
contribution to it, taken from fig.~\ref{fig:tjbrw}. A not very
large deviation from the SM value for this cross section would
therefore imply the presence of sizeable FCNC contributions, even
with a measurement of $Br(t\,\rightarrow\,b\,W)$ in full agreement
with its expected SM values.

Finally, we may use the current values for
$Br(t\,\rightarrow\,b\,W)$, assuming unitarity of the CKM matrix, to
ask: what is the maximum value for the FCNC branching ratios that is
allowed under those assumptions? The answer lies in
fig.~\ref{fig:qXbrw}, where we plot the sum of all top FCNC
branching ratios versus $1\,-\,Br(t\,\rightarrow\,b\,W)$. The data
shown in this plot already takes the Tevatron results
on single top production into account as well as the constraints from B-physics.
Results from the Tevatron are very important but the results from B-physics
are even more restrictive and this is in full display
in fig.~\ref{fig:brWbrwe}. In this figure we plotted
$1\,-\,Br(t\,\rightarrow\,b\,W)$ against the electroweak FCNC
branching ratios - we can appreciate just how much parameter space
is excluded by the B-physics results from~\cite{Fox:2007in}.


Reading the intersection between the SM-predicted value for
$1\,-\,Br(t\,\rightarrow\,b\,W)$ and the band of values allowed for
the FCNC branching ratios from fig.~\ref{fig:qXbrw}, we obtain the
following upper bound for the total FCNC branching ratio:
\begin{equation}
Br(t\,\rightarrow\,q\,X)\;<\; 4.0\times 10^{-4} \;\;\;
\label{eq:bound}
\end{equation}
while using only Tevatron data the constraint would be $1.0\times 10^{-3}$,
again showing the relevance of the B-physics results.
If we now read the intersection between the SM-predicted value for
$1\,-\,Br(t\,\rightarrow\,b\,W)$ and the band of values allowed for
the FCNC branching ratios from fig.~\ref{fig:brWbrwe}, we obtain a very similar
upper bound for the electroweak FCNC branching ratio:
\begin{equation}
Br(t\,\rightarrow\,q\,Z) + Br(t\,\rightarrow\,q\,\gamma)\;<\; 4.0\times 10^{-4} \;\;\;.
\label{eq:bound}
\end{equation}
The best current upper bound for the electroweak sector are 
the Hera and LEP for $Br(t\, \rightarrow \,q \, \gamma)$ and from
the Tevatron for $Br(t\, \rightarrow \,q \, Z)$ an are all at
the percent level. Therefore our results are better by at least
one order of magnitude. Moreover they are obtained for the
sum of branching ratios with $q=u,c$. Therefore they apply
to each individual branching ratio. 

The excellent results obtained by the Tevatron from direct top
production~\cite{Aaltonen:2008qr} agree with ours for $q=u$ and our
results are slightly better for $q=c$. We remind the reader that our results were obtained
under one important theoretical assumption namely,
the unitarity of the CKM matrix. 


\section{Conclusions}
\label{sec:conc}

We considered a vast set of dimension six effective operators
contributing to FCNC interactions of the top quark. In previous
works the effects of those operators in processes of single top
production was analysed in great detail. In this letter, we
concerned ourselves with a more basic observable, the main branching
ratio of the top, $Br(t\,\rightarrow\,b\,W)$. Several of the
operators that contribute to FCNC decays of the top also affect its
charged decays. We thus have contributions from FCNC anomalous
couplings to top charged branching ratios, through interference
terms between SM Feynman diagrams and diagrams containing an
anomalous vertex. Those FCNC contributions would then constitute
corrections to the top quark CKM matrix elements.

We computed $Br(t\,\rightarrow\,b\,W)$ for a vast set of possible
values for the FCNC anomalous couplings, and analysed the cross
sections for several processes of single top production via FCNC at
the LHC. The values predicted by the SM for
$Br(t\,\rightarrow\,b\,W)$, if one assumes unitarity of the CKM
matrix, are very precise and give us an immediate upper bound for
the FCNC cross sections. A measurement of a LHC cross section for
$t\,+\,Z$ production at the LHC superior to about 1 pb, for
instance, would imply that the CKM matrix is in fact non-unitary.

Assuming unitarity of the CKM matrix also allows us to obtain a
rather good upper bound on the total FCNC branching ratio, better
than most current direct experimental measurements for operators 
in the electroweak sector. For operators stemming from
the strong sector, the CDF results are competitive with ours
especially the ones involving the $u$-quark. The
curious interplay between top FCNC interactions and its charged
decays, presented in this paper, allows a consistency check between
several top quarks observables. If the top quark indeed has sizeable
FCNC interactions, stemming from some type of new physics, the
correlations between these several observables will provide us with
a better understanding of those interactions.

\vspace{0.25cm} {\bf Acknowledgments:} Our thanks to Veronique
Boisvert, for many useful discussions and for a careful reading of the manuscript. This work is supported by
Funda\c{c}\~ao para a Ci\^encia e Tecnologia under contract
PTDC/FIS/70156/2006. R.S. is supported by the FP7 via a Marie
Curie Intra European Fellowship, contract number
PIEF-GA-2008-221707.

\end{document}